\documentclass[a4paper,11pt]{article}
\usepackage[left=2.5cm,top=3cm,right=2.5cm,bottom=3cm,bindingoffset=0.0cm]{geometry}

\usepackage[utf8]{inputenc}  % Allow UTF-8 encoded characters
\usepackage[T1]{fontenc}     % Better accented characters
\usepackage{amsmath, amssymb}
\usepackage{graphicx}        % Include figures
\usepackage{hyperref}        % Hyperlinks in PDF
\usepackage{booktabs} % Enhances table formatting with better rules (lines)
\usepackage{siunitx}
\usepackage{subcaption}
\usepackage[backend=biber,
            style=numeric,
            datamodel=standard,
            sorting=none,
            doi=true,
            url=false,
            isbn=false]{biblatex}
\usepackage{color}
\usepackage{multirow}
\usepackage[rightcaption]{sidecap} % put caption on the right; use [leftcaption] for left

\sidecaptionvpos{figure}{c} % or {b}

\title{Stochastic Point Kinetics Model of Circulating-Fuel Reactors under Perfect Mixing Approximation}

\usepackage{authblk}
\author[1]{Lubomír Bureš}
\author[2]{Valeria Raffuzzi}
\affil[1]{Saltfoss Energy ApS, Titangade 11, 2200 Copenhagen, Denmark \vspace{1ex}}
\affil[2]{Department of Engineering, University of Cambridge, Cambridge CB2 1PZ, United Kingdom \vspace{1ex}}
\affil[ ]{$^1$\texttt{lubomir.bures@saltfoss.com}, $^2$\texttt{vr339@cam.ac.uk}}
\date{January 21, 2026}

\begin{document}
\maketitle

\begin{abstract}
  \noindent We present a stochastic framework for low-population dynamics in circulating-fuel reactors (CFRs) that captures delayed-neutron precursor (DNP) transport without delay terms. Starting from a modified point-kinetics model with two perfectly-mixed volumes, we derive equivalent discrete-event dynamics and an Itô stochastic differential equation (SDE) system. Two solvers are implemented: an analog Monte Carlo (AMC) engine and a semi-implicit Milstein SDE solver. Transient benchmarks demonstrate perfect agreement of AMC/SDE means with deterministic solutions, while revealing that the SDE approach underestimates DNP variances in selected regimes, potentially due to the neglect of DNP noise. We further recast reactivity loss due to precursor drift in this stochastic setting and show that its estimator is negatively biased. Overall, the developed framework provides a minimal yet representative model for CFR low-population kinetics. Future work will re-derive and test SDE noise terms and apply the framework to selected transient applications such as start-up analyses of CFRs.

\end{abstract}

\section{Introduction}
\label{sec:intro}
When studying the low-population dynamics of nuclear reactors, two approaches have been prominently featured in previous work: probability-balance equations, such as the Pál-Bell formalism \cite{Chang2022,Gordon2021}, and the stochastic-point-kinetic approach proposed in \cite{Hayes2005} using stochastic differential equations (SDEs). The latter approach was recently re-derived in \cite{Bonnet2025} using a martingale representation in the diffusive approximation and then benchmarked against reference analog Monte Carlo (AMC) solutions using a set of model problems. The interested reader is referred to \cite{Gordon2021} for a comprehensive review of modelling approaches.

\begin{SCfigure}
  \includegraphics[width=.4\linewidth]{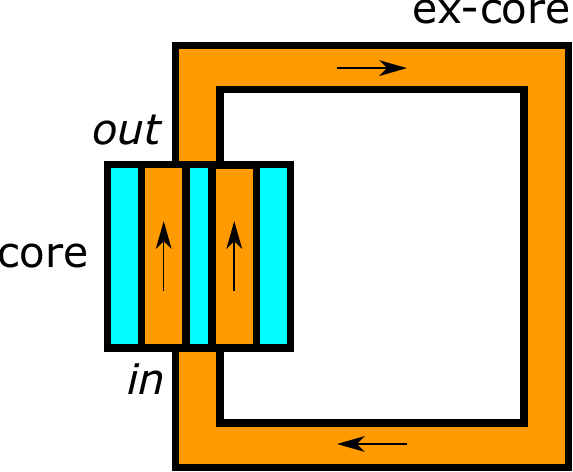}
  \caption{Schematic layout of the system considered in this work. The core region \textbf{\textit{c}} is indicated with the rest of the system being labelled as ex-core \textbf{\textit{e}}. Inlet and outlet locations of the core are indicated and arrows are included to show the direction of the fuel flow.}
  \label{fig:cstr}
\end{SCfigure}

To the best of authors' knowledge, low-population dynamics studies have been limited to static-fuel reactors so far. For circulating-fuel reactors (CFRs), the advection of delayed neutron precursors (DNPs) with the flow of the fuel represents a key challenge; to demonstrate this, consider now the simple embodiment of a CFR shown in Fig.\ \ref{fig:cstr} consisting of two control volumes, the core region $c$ and the ex-core region $e$, which are connected together to form a loop, i.e.\ the inlet of one of the regions is the outlet of the other and vice versa. After the first attempts at developing a dynamic model of such a system, notably by Ergen \cite{Ergen1954} and Fleck \cite{Fleck1955}, MacPhee \cite{Macphee1958} proposed the modified point kinetics in the well-known form used until the present day (with source term added):
\begin{align}
	\frac{dN(t)}{dt}&=\frac{\rho(t)-\beta}{\Lambda}N(t)+\sum_j\lambda_j C_j(t) + S(t), \label{eq:pow_pk_cstr}\\
	\frac{d C_j(t)}{d t}&=\frac{\beta_j}{\Lambda} N(t)-\lambda_j C_j(t) + \frac{1}{\tau_c}\Big(C_j(t-\tau_e)\exp(-\lambda_j \tau_e)-C_j(t)\Big). \label{eq:prec_pk_cstr}
\end{align}
Here, $t$ (s) is time, $N(t)$ (-) the neutron population, $\beta$ (-) the total delayed neutron fraction, $\Lambda$ (s) the prompt neutron generation time, $\rho(t)$ (-) the reactivity, $\lambda_j$ (1/s) the decay constant of the precursor group $j$, $C_j(t)$ (-) the population of precursors in the active core, the importance being neglected, and $S(t)$ (1/s) the external source rate. Additionally, $\tau_c$ and $\tau_e$ (s) are the precursor residence times  in the core and ex-core regions, which are taken to be constant. An examination of the above equations reveals that the presence of a delay term in \eqref{eq:prec_pk_cstr} introduces a ``memory'' effect into the problem. As a result, an equivalent Markov process for low-population dynamics cannot be found, since for a continuous-time Markov process, the future evolution of the system must depend only on the current state \cite{Lawler2014}.

In principle, this problem could be approached by replacing the delay term, which corresponds to a plug-flow representation of the ex-core region, by an explicit treatment of the precursor flow in the form of an advection-reaction equation, elaborated on, e.g.,\ in~\cite{Morgan2018,Bures2024}. However, it was observed in \cite{Bures2025} observed that the plug-flow representation of the ex-core region is only one of many options and, in some cases, not necessarily the most accurate one due to the potential presence of strong mixing. For this reason, it is for example possible to opt for the perfectly-mixed approximation in both core and ex-core regions. The resulting modified point kinetics equations then take the form:
\begin{align}
	\frac{dN(t)}{dt}&=\frac{\rho(t)-\beta}{\Lambda}N(t)+\sum_j\lambda_j C_{j,c}(t) + S(t), \label{eq:mpk_power}\\
	\frac{d C_{j,c}(t)}{d t}&=\frac{\beta_j}{\Lambda} N(t)-\lambda_j C_{j,c}(t)-\frac{C_{j,c}(t)}{\tau_c} + \frac{C_{j,e}(t)}{\tau_e}, \label{eq:mpk_prec_c}\\
	\frac{d C_{j,e}(t)}{d t}&=-\lambda_j C_{j,e}(t)-\frac{C_{j,e}(t)}{\tau_e} + \frac{C_{j,c}(t)}{\tau_c}. \label{eq:mpk_prec_e}
\end{align}
Here, subscripts $c$ and $e$ are used to indicate the distinct precursor populations in the two regions of the system. The above set of equations retains the simplicity of point kinetics and does not feature any delay terms at the cost of doubling the number of equations needed for tracking the DNP population. Despite not being appropriate in all circumstances, the simple framework of two periodically connected perfectly-mixed volumes is considered to be sufficient for the purpose of the present work, i.e.,\ the first attempt at developing models for studying the low-population dynamics of CFRs. 

The rest of the paper is organised as follows: in Section \ref{sec:theor}, the theoretical framework used in this work is presented. Then, in Section \ref{sec:implement}, the corresponding numerical implementation is briefly discussed for AMC and SDE approaches. In Section \ref{sec:bench}, the two numerical approaches are benchmarked against one another and, in Section \ref{sec:appl}, the methodology is showcased using a reactivity loss calculation. Finally, conclusions are drawn in Section \ref{sec:conc}.

\section{Theoretical model} 
\label{sec:theor}

We start by formulating the dynamics of a continuous-time stochastic process equivalent to the point-kinetic model given in the previous section (Eqs.\ \ref{eq:mpk_power}, \ref{eq:mpk_prec_c}, and \ref{eq:mpk_prec_e}). For simplification purposes, only one group of precursors is considered with the extension to multiple groups following naturally. The compiled dynamics are shown in Table \ref{tab:dynamics_i}. In this table, $(n_t,\,c_{c,t},\,c_{e,t})_{t\geq 0}$ take discrete values and describe neutron and DNP population sizes, the latter being considered in both regions of the system. Furthermore, neutrons in this process are considered to be lost either through capture or leakage with per-particle rate $\gamma(t)$ and to induce fission with rate $\phi$. The production of prompt neutrons and DNPs due to fission, i.e.\ the fission yield, is modelled as two random variables $\nu_p$ (prompt) and $\nu_d$ (delayed) described by Poisson distributions  with averages $\mathbb{E}[\nu_p]$ and $\mathbb{E}[\nu_d]$, respectively. The external source of neutrons is considered to be modelled with a Poisson distribution with intensity $S(t)$. Precursors are transferring between the two regions of the system with rates given by the respective residence times and DNP decay produces a neutron only if the decay occurs in the core.

The fission and loss rates per neutron $\phi$ and $\gamma(t)$ can be expressed with the help of the variables and parameters introduced in the previous paragraph as \cite{Sutton2017}:
\begin{align}
    \phi &= \frac{1 - \beta}{\mathbb{E}[\nu_p] \cdot \Lambda}, \label{eq:phi}\\
    \gamma(t) &= \frac{1 - \rho(t)}{\Lambda} - \phi.
\end{align}
Note that, in principle, even the fission rate $\phi$ could be time-dependent. However, in the present point-kinetic formulation, it is computed from constants \cite{Akcasu1971} and, thus, itself a constant.

\begin{table}[ht]
    \centering
	\caption{Dynamics of the modelled stochastic process.}
    \begin{tabular}{l|l|l}
        \hline
        \textbf{Event} & \textbf{Dynamics} & \textbf{Rate} \\ \hline
        Loss (capture \& leakage)     & $(n_t,\,c_{c,t},\,c_{e,t}) \;\rightarrow\; (n_t - 1,\, c_{c,t},\, c_{e,t})$                       & $n_t\cdot \gamma(t)$        \\[4pt]
        Fission      & $(n_t,\,c_{c,t},\,c_{e,t}) \;\rightarrow\; (n_t + \nu_p - 1,\, c_{c,t} + \nu_d,\, c_{e,t})$      & $n_t\cdot \phi$         \\[4pt]
        External source  & $(n_t,\,c_{c,t},\,c_{e,t}) \;\rightarrow\; (n_t + 1,\, c_{c,t},\, c_{e,t})$                        & $S(t)$                  \\[4pt]
        DNP decay in core       & $(n_t,\,c_{c,t},\,c_{e,t}) \;\rightarrow\; (n_t + 1,\, c_{c,t} - 1,\, c_{e,t})$                    & $c_{c,t}\cdot \lambda$       \\[4pt] 
	    DNP decay in ex-core       & $(n_t,\,c_{c,t},\,c_{e,t}) \;\rightarrow\; (n_t,\, c_{c,t},\, c_{e,t} - 1)$                    & $c_{e,t}\cdot\lambda$       \\[4pt] 
        Core $\rightarrow$ ex-core DNP transfer       & $(n_t,\,c_{c,t},\,c_{e,t}) \;\rightarrow\; (n_t,\, c_{c,t} - 1,\, c_{e,t} + 1$)                    & $c_{c,t}/\tau_c$       \\[4pt] 
        Ex-core $\rightarrow$ core DNP transfer       & $(n_t,\,c_{c,t},\,c_{e,t}) \;\rightarrow\; (n_t,\, c_{c,t} + 1,\, c_{e,t} - 1)$                 & $c_{e,t}/\tau_e$       \\[4pt] 
		\hline
    \end{tabular}
	\label{tab:dynamics_i}
\end{table}

Upon reviewing the derivation of the Itô process in the case of static-fuel reactors presented in \cite{Bonnet2025} and considering the current situation, i.e.\ discrete dynamics given in Table \ref{tab:dynamics_i} and deterministic point-kinetic equations \eqref{eq:mpk_power}, \eqref{eq:mpk_prec_c}, and \eqref{eq:mpk_prec_e}, one can derive in a similar manner that the system of SDEs describing the evolution of neutron and DNP populations in the system considered in this work is given as:
\begin{align}
	dN_t &= \Bigg([\rho(t)-\beta]/\Lambda\cdot N_t + \sum_j \lambda_j \cdot C_{j,c,t} + S(t)\Bigg)dt + D(t)\sqrt{|N_t|}\,dW_t.  \label{eq:sde_power} \\
	dC_{j,c,t} &= \Bigg(\beta_j/\Lambda\cdot N_t -(\lambda_j+1/\tau_c) \cdot  C_{j,c,t} + 1/\tau_e\cdot C_{j,e,t}\Bigg)dt,  \label{eq:sde_prec_c}\\
	dC_{j,e,t} &= \Bigg(-(\lambda_j+1/\tau_e) \cdot  C_{j,e,t} + 1/\tau_c\cdot C_{j,c,t}\Bigg)dt. \label{eq:sde_prec_e}
\end{align}
In the above equations, $dW_t$ is a regular Brownian motion and the assumption of single DNP group was relaxed. As also noted in \cite{Bonnet2025}, from a numerical point of view nothing prevents $(N_t,\,C_{c,t},\,C_{e,t})$ from attaining negative values. This leads to the need to take the absolute value in the diffusion term. Note also that the diffusion coefficient $D(t)$ can be found to be equal to:
\begin{equation}
     D(t) = \sqrt{\mathbb{E}[(\nu_p - 1)^2] \cdot \phi + \gamma(t)}.
\end{equation}

\section{Numerical implementation}
\label{sec:implement}
In this work, we consider two modelling approaches: direct simulation of the stochastic process described in Table \ref{tab:dynamics_i} via analog Monte Carlo (AMC) simulation and solution of the system of SDEs given by Eqs.\ \ref{eq:sde_power}, \ref{eq:sde_prec_c}, and \ref{eq:sde_prec_e}. Both are briefly described in this section.

\subsection{Analog Monte Carlo solver MARS}
\label{sec:mars}
MARS is an AMC stochastic point kinetics solver, designed after the solver Marduk~\cite{Sutton2017}. MARS models the evolution of the neutron and precursor populations by explicitly simulating the events described in Table~\ref{tab:dynamics_i}, with the difference that it also discriminates between prompt and delayed neutrons. For this paper, the capability to track ex-core DNPs was added compared to Marduk's capabilities. MARS simulates the given transient a number of times chosen by the user, and collects statistics over these independent replicas. This loop can be parallelised with MPI. Uniform or linear time profiles of reactivity and precursor residence times are allowed.

\subsection{Itô Process (SDE) solver}
\label{sec:sde}
The governing equations \eqref{eq:sde_power}, \eqref{eq:sde_prec_c}, and \eqref{eq:sde_prec_e} were discretised in time using the semi-implicit Milstein method, which was  implemented in Python 3.12 using the NumPy library. Non-uniform time step informed by an initial solution of the deterministic equations with a 5th-order Radau IIA solver from the SciPy library was used. If we label $\mathbf{X}^n = [N(t_n), \mathbf{C_c}(t_n), \mathbf{C_e}(t_n)]^\mathsf{T}$, advancing the solution in time corresponds to the following two-step procedure:
\begin{enumerate}
    \item \textit{Deterministic drift}: $\mathbf{X}^* = \mathbb{M}^n\mathbf{X}^n + \mathbf{m}^n$.
    \item \textit{Stochastic increment with Milstein correction}: $\mathbf{X}^{n+1} = \mathbf{X}^* + \mathbf{K}^n$.
\end{enumerate}
In the above:
\begin{align}
     \mathbb{M}^n &= [\mathbb{I} - \Delta t_n \mathbb{A}(t_{n+1})]^{-1},\\
     \mathbf{m}^n &= \Delta t_n \mathbb{M}^n \mathbf{b}(t_{n+1}),\\
     K^n_0 &= D(t_n)\sqrt{\mathbf{X}_0^n} \Delta W_n + 0.25\,D(t_n)^2\big(\Delta W_n^2 - \Delta t_n\big).
\end{align}
with $\mathbb{A}$ being the system matrix, $\mathbf{b}$ the right-hand side vector, $\Delta t_n$ the time step length, and $\Delta W_n$ the normally-distributed stochastic increment with zero mean and variance $\Delta t_n$. Only the zeroth component of $\mathbf{K}$, i.e.\ $K_0$, corresponding to the noise in neutron population is non-zero.

To facilitate an efficient evaluation of the ensemble of solutions needed for statistics acquisition, the terms $\mathbb{M}$, $\mathbf{m}$, and $D(t_n)$ are pre-computed for all time steps. This allows for the use of efficient and scalable general matrix multiplies, traversing solution paths in a vectorised manner.

\section{Benchmarking} 
\label{sec:bench}

To benchmark the two approaches presented above, we simulated several cases inspired by the ``ramp reactivity insertion problem" from \cite{Sutton2017}. The parameters used are given in Table \ref{tab:bench}. Note that the neutron generation time was increased from the value given in \cite{Sutton2017} in order to satisfy the source positivity condition $S(t)>D(t)^2/2$, needed to guarantee that the sample trajectories of the modelled Itô process are almost surely strictly positive \cite{Bonnet2025}. In our preliminary test calculations we confirmed the observations from \cite{Bonnet2025}, namely that allowing negative trajectories would lead to over-estimation of calculated variances and that clamping of trajectories would lead to introducing bias in the mean values. Thus, in order to successfully simulate the Itô process with the presented numerical model, the source positivity condition must be satisfied.

\begin{table}[ht]
\centering
\setlength{\tabcolsep}{6pt}
\renewcommand{\arraystretch}{1.15}

\caption{Problem parameters for the calculation cases.}
\begin{tabular}{@{} l l *{6}{c} @{}}
\toprule
\textbf{Global parameters}
  & \multicolumn{7}{l}{$\Lambda=10^{-3}\ \mathrm{s},\; S=8800\ \mathrm{n/s},\; \beta=0.0065$,\; $[N(0), \mathbf{C_c}(0), \mathbf{C_e}(0)]^\mathsf{T}=\mathbf{0}$} \\
\midrule

\cmidrule(l){3-8}
\multicolumn{2}{l}{} % empty label + empty sublabel column
  Group number & \textbf{1} & \textbf{2} & \textbf{3} & \textbf{4} & \textbf{5} & \textbf{6} \\
\multirow{2}{*}{\textbf{DNP data}}
  & $\alpha_i = \beta_i/\beta$ (-) & 0.033  & 0.219  & 0.196  & 0.395  & 0.115 & 0.042 \\
  & $\lambda_i\;(\mathrm{s}^{-1})$ & 0.0124 & 0.0305 & 0.111  & 0.301  & 1.14  & 3.01 \\
\cmidrule(l){3-8}
\addlinespace[3pt]
\multirow{2}{*}{\textbf{Distribution of $\nu_p$}}
  & Multiplicity (-)     & 0 & 1 & 2 & 3 & 4 & 5 \\
  & Abundance (-) & 0.027 & 0.158 & 0.339 & 0.305 & 0.133 & 0.038 \\
\bottomrule
\end{tabular}

\label{tab:bench}
\end{table}

Several transients were considered in this exercise. Here we highlight the ``ramp-up'' problem defined by the schedules:
    \begin{align}
        \rho(t) &= -0.01 \cdot [1 - \min(t/10,1)] + 0.006 \cdot \min(t/10,1),\\
        \tau_c(t) &= 1000 \cdot  [1 - \min(t/5,1)] + 10 \cdot \min(t/5,1),\\
        \tau_e(t) &= 1000 \cdot [1 - \min(t/5,1)]  + 15 \cdot  \min(t/5,1).
    \end{align}
By including simultaneous variations of both reactivity and residence times, a case such as this one should be able to ``stress-test'' the computational approaches. Nevertheless, in an actual application, the residence times generally ought to be kept constant to guarantee the validity of the assumptions underlying the physical modelling. 

\begin{figure}
  \centering
  \begin{subfigure}{0.47\textwidth}
    \centering
    \includegraphics[width=\linewidth]{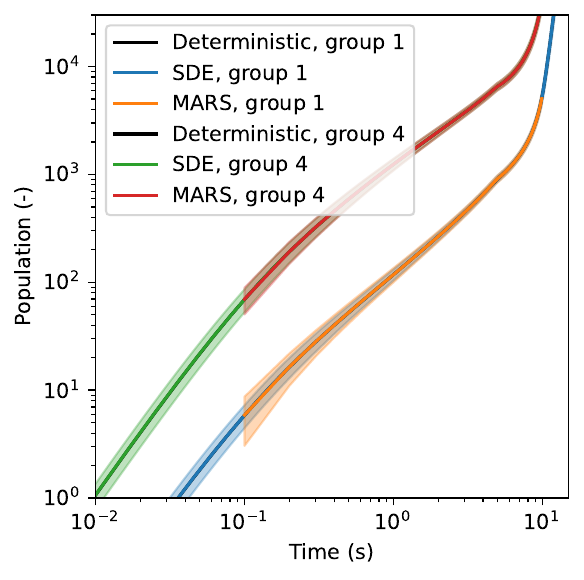}
    \caption{In-core population.}
    \label{fig:rampup_cc}
  \end{subfigure}\hfill
  \begin{subfigure}{0.47\textwidth}
    \centering
    \includegraphics[width=\linewidth]{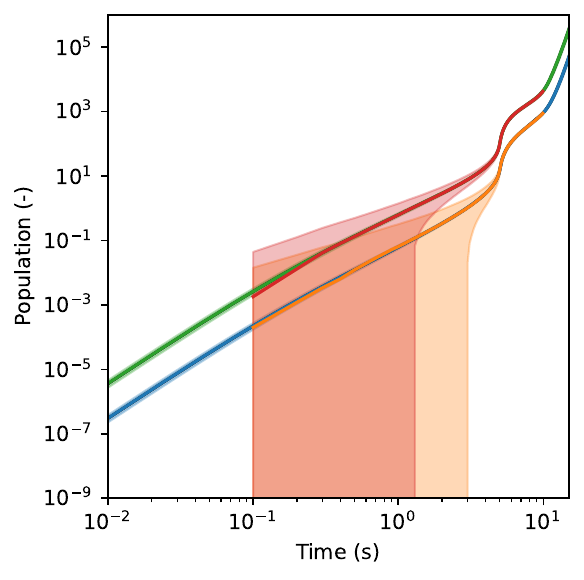}
    \caption{Ex-core population.}
    \label{fig:rampup_ce}
  \end{subfigure}
  \caption{Population of selected precursor groups in the ``ramp-up'' transient benchmark case. Shaded regions correspond to $\pm1\sigma$. The deterministic solution is not visible due to perfect overlap with the mean trajectories.}
  \label{fig:rampup}
\end{figure}

For the SDE solver, 8000 trajectories were simulated to obtain converged statistics. For MARS, as little as 100 independent repetitions were enough to convergence all statistics; however, \num{10000} repetitions were used for better precision. The mean values agreed perfectly between the two methods and when compared to the deterministic solution of the modified point kinetics equations. This is illustrated for DNP groups 1 and 4 in Fig.\ \ref{fig:rampup}. While for large neutron and DNP populations the variance was found to agree between MARS and the SDE solver as well, in some cases, such as for the in-core group 1 and ex-core groups shown in Fig.\ \ref{fig:rampup}, it can be seen that the variance obtained by the SDE solver is lower than the one predicted by MARS. This was confirmed for simulations of different transients, transients with different initial conditions, and also for steady-state problems with and without flow. Thus, this discrepancy is considered to be a function of the solution method rather than of the simulated case in question. We speculate that this is due to the fact that, in the SDE approach, the noise in DNP evolution is neglected \cite{Bonnet2025}. We note that the equations describing the Itô process presented in \cite{Bonnet2025} do include the DNP noise term, which is only neglected by choice afterwards. However, in our re-derivation of the equations, we found that the DNP noise disappears in the diffusion limit identically due to the thinning of the production of delayed neutrons in the sequence of rescaled models introduced to keep the precursor deterministic drift finite. Thus, we suggest that the derivation of the governing SDEs should be scrutinised in future work, especially since the elimination of DNP noise appears to lead to an under-estimation of the DNP population variance in selected configurations.

\section{Reactivity loss calculation} 
\label{sec:appl}
\begin{SCfigure}
  \includegraphics[width=.6\linewidth]{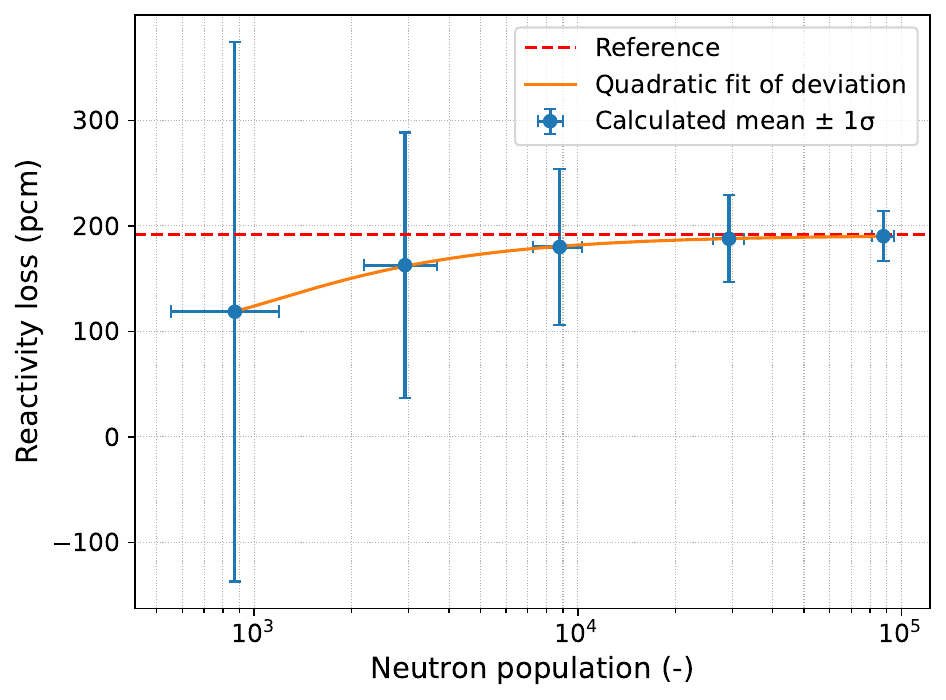}
  \caption{Estimated reactivity loss due to DNP drift $\mathbb{E}[\hat{\rho}_0]$ as a function of steady-state neutron population. The reference value $\rho_0 = 191.5$ pcm as well as the quadratic fit of $(\mathbb{E}[\hat{\rho}_0]-\rho_0)$ as a function of $1/N_0$ are also shown.}
  \label{fig:rho0}
\end{SCfigure}

Stochastic methodologies are typically used for safe start-up studies \cite{Gordon2024}. However, a definition of a safe start-up requires firstly a definition of undesired stochastic transients. For non-conventional reactors with circulating fuel, such as molten-salt reactors, this is a non-trivial task, the conclusion of which will likely depend on the particular design in question as well as on the regulatory framework employed for the licensing of the given reactor. Thus, we do not attempt to delve into this topic in this paper and defer it to future work. Instead, we examine the widely studied reactivity loss due to precursor drift from the novel angle of low-population dynamics. The interested reader is referred to \cite{Bures2025} for extensive discussion of this quantity, which, in the current set-up, can be derived to be:
\begin{equation}
	\rho_0 = \beta - \frac{\Lambda}{N_0} \cdot \sum_j\lambda_j\cdot C_{j,c,0}.
\end{equation}
This equation can be shown to be valid under both critical and sub-critical steady-state conditions. As one can see, $\rho_0$ is a non-linear function of $N$ and $\mathbf{C_c}$. As a result, when tallying this quantity on-the-fly, we expect it to be, by Jensen's inequality, negatively biased with respect to the deterministic value. For the configuration described in Table \ref{sec:bench} with $\tau_c = $\SI{10}{s} and $\tau_e = $ \SI{15}{s} under varying degrees of subcriticality, we obtained the result shown in Fig.\ \ref{fig:rho0}, indeed showing the negative bias of estimated reactivity loss. Due to the long times needed for steady-state convergence, only the SDE solver with 8000 trajectories was used. Thus the variance of $\hat{\rho}_0$ is potentially under-estimated as discussed in the previous section. This effect is however considered to be minor, since the populations of DNPs were in general significantly higher than those in the benchmarking cases. 

We believe the bias in $\mathbb{E}[\hat{\rho}_0]$ to be worth highlighting, as it does not appear to have been discussed in the open literature, despite the general bias in estimators involving ratios of correlated averages being well-known. This observation could be relevant for the development of estimation methods for $\rho_0$ in time-dependent Monte Carlo models, which is a topic of active research \cite{Kim2022}.

We conclude this discussion by noting that, as seen in Fig.\ \ref{fig:rho0}, $\mathbb{E}[\hat{\rho}_0]$ converges to $\rho_0$ as $\mathcal{O}(1/N_0^2)$. This is expected: to first order in fluctuations, $\chi =\sum_j\lambda_j C_{j,c}$ is a linear time-invariant low-pass filter of $N$ with DC gain $r_0=\chi_0/N_0$. Thus, the leading $\mathcal{O}(1/N_0)$ order variance and covariance terms in the second-order delta-method expansion of $\mathbb{E}[\hat{r}]$ with $r=\chi/N$ cancel out exactly and the convergence of $\mathbb{E}[\hat{\rho}_0]$ is driven by the higher-order deviation of the filter from its DC gain. For the same reason, $\text{Var}(\hat{\rho}_0)$ behaves roughly like $\mathcal{O}(1/N_0)$. However, $\text{Var}(\hat{N}_0)$ appears to behave like $\mathcal{O}(N_0^{4/3})$, which deviates from the expected diffusion-like $\mathcal{O}(N_0)$. In future work, we will investigate whether this is a function of the numerical treatment of the governing equations or actual physical effects.

\section{Conclusions}
\label{sec:conc}
We developed a stochastic framework for low-population dynamics in circulating-fuel reactors by casting a modified point-kinetics model featuring two perfectly-mixed volumes into equivalent discrete-event dynamics and Itô SDEs, and implemented both an AMC solver and a Milstein SDE solver. Benchmarks showed perfect agreement of mean trajectories with deterministic solutions, while the SDE approach underestimates DNP variances in some regimes, motivating a closer examination of the derivation, particularly the treatment of DNP noise. Recasting reactivity loss due to DNP drift in this stochastic setting revealed a negatively biased estimator, a result potentially pertinent to time-dependent Monte Carlo estimations. Future work will re-derive and test SDE noise terms, examine the convergence behaviour of $\text{Var}(\hat{N}_0)$, and apply the developed framework to transient applications such as safe start-up analyses of circulating-fuel reactors.

\section*{Declaration of competing interest}
Lubomír Bureš declares a competing interest due to affiliation with Saltfoss Energy ApS, a company actively involved in the development of circulating-fuel, molten-salt reactor technology, in the form of employment.

\section*{Acknowledgements}
Valeria Raffuzzi was supported by the EPSRC grant MaThRad (EP/W026899/1). She acknowledges Theophile Bonnet for useful discussions.

%------ References ------
%\clearpage
\printbibliography

\end{document}